\documentclass[a4paper,11pt]{article}
\usepackage{jheppub} 
\usepackage{lineno}
\usepackage{url}
\usepackage{subcaption} 

 \newcommand{\Ord}[2]{\mathcal O \left(#1\right)^{#2}}

\arxivnumber{ } 

\title{\boldmath Gregory-Laflamme-type instability of boson strings and related phases  
in $D=5$ Kaluza-Klein theory }

\author{C. Herdeiro} 
\author{and E. Radu} 
\affiliation{Departamento de Matem\'atica  da Universidade de Aveiro and
	Center for Research and Development in Mathematics and Applications (CIDMA)
	\\Campus de Santiago, 3810-183 Aveiro, Portugal}

\abstract{
We add an $S^1$ extra dimension (size $L$) to the well known $D=4$ static, spherically symmetric
$Q$-balls and boson stars.
We show that the resulting {\it uniform horizonless boson strings} possess a static zero-mode 
for a critical value of $L$. This is at the threshold of a Gregory-Laflamme instability of these objects, occuring for larger values of $L$. 
 The non-linear continuation of the zero-mode yields $D=5$ {\it non-uniform boson strings}.
 In addition, there are also intrinsic $D=5$
 solutions describing {\it localized boson stars on the $S^1$}, supported by the contribution
of the scalar field Kaluza-Klein modes. 
 Basic properties of all three types of aforementioned solutions are discussed, 
 together with their phase space.

}

\begin{document}
\maketitle
\flushbottom

 \section{ Introduction and motivation}

The dimensionality of space (time) is a long-standing
fundamental question in physics and mathematics,
which goes back to Aristotle \cite{1}.
In a modern context, this question was first raised a century ago in the pioneering work of 
Nordstr\"om \cite{Nordstrom:1914ejq}, 
Kaluza \cite{Kaluza}
and 
Klein \cite{Klein:1926tv}.
Attempting the unification of the
then known interactions, namely gravity and electromagnetism,
they assumed the existence of four spatial dimensions;
however, one dimension should form a circle so small as to be unobservable.
Then (vacuum) gravity in
five (spacetime) dimensions,
as described by  (extrapolating) Einstein’s theory of General Relativity (GR), has the appearance, along a four dimensional ``slice", of
Einstein-Maxwell theory
with an extra scalar field, often called \textit{dilaton}.  

This idea has proven one of the most fruitful
in theoretical physics,
the original Kaluza-Klein (KK) model being extended in various directions,
such as considering
more than one extra dimension
and/or the inclusion of matter fields
(see~\cite{ACF} for a review and a large set of original references). 
In an attempt to provide a unified framework for all interactions,
string/M-theory  predicts the existence of compactified extra dimensions, reviving the KK idea in the last half century. 

\medskip

Consider the original KK framework
with a single $S^1$ compact extra-dimension (size $L$), a four dimensional non-compact manifold ${\cal M}_4$
and configurations with standard ${\cal M}_4\times S^1$ asymptotics.
Then, one finds $three$ different classes of 
(static, vacuum)
solutions of 
Einstein's equations  
(see~\cite{Kol:2004ww,Harmark:2007md};
in what follows, we shall exclude KK bubbles):

\begin{itemize}
\item 
{\bf Uniform black strings} - 
The simplest solution is obtained by adding a trivial extra dimension to the $D=4$ Schwarzschild black hole, with horizon radius $r_H$. 
It is translational invariant along the extra-coordinate and describes a \textit{uniform black string}, with an event horizon topology $S^2 \times S^1$. But while the $D=4$ Schwarzschild
black hole  is stable against perturbations, the black string is linearly unstable  against long wavelength perturbations  along the $S^1$-circle, 
as found by Gregory and Laflamme (GL)
\cite{Gregory:1993vy}.
%
%
That is, for a given $r_H$,
there is a critical value of $L$, denoted  $L_0$, of order $r_H$,
such that 
 black strings with
 $L<  L_0$ ($L> L_0$ ) are stable (unstable).
At the threshold, $L=L_0$, there
is a static zero mode perturbations, breaking the translational invariance of the string along the circle. This is a linear manifestation of the bifurcation to a new phase (next bullet point).

\item 
{\bf Non-uniform black strings} - Emerging smoothly from the uniform string at the critical point where stability changes, a family of \textit{non-uniform black strings} are the non-linear continuation of the aforementioned zero mode.
They were first
found perturbatively from the critical GL string 
\cite{Gubser:2001ac,Wiseman:2002zc,Sorkin:2004qq},
being later numerically extended
into the full nonlinear regime\footnote{See also the pioneering work 
\cite{Wiseman:2002zc},
which reports $D=6$ results. }
in~\cite{Kleihaus:2006ee}.
The horizon topology
of these solutions
is still $S^2\times S^1$, but
their geometry has a nontrivial dependence along the extra dimension.
  
\item 
{\bf Localized black holes} - A third set of solutions, bearing relevance to the dynamical development of the GL instability of the uniform strings, have an event horizon of $S^3$-topology,
despite still possessing ${\cal M}_4\times S^1$ asymptotics
\cite{Kudoh:2004hs}.
The (hyper)spherical $D=5$ Schwarzschild-Tangherlini BH with horizon size $r_H$
provides a good approximation 
for $r_H \ll L$.
As their horizon size (and mass) grows with respect to $L$,
the horizon deforms to a prolate ellipsoid.
 
\end{itemize}

These three classes of solutions turn out to combine in a complicated phase diagram,
with some surprising features.\footnote{
No explicit solutions in closed analytical form are available for non-uniform black strings
and (vacuum) localized black holes.
However, one
can analyse their properties by using a combination of  numerical and  analytical  methods,
which is enough for most purposes.
}

The GL instability can be understood heuristically by using an entropy argument: for a (uniform) black string horizon with fixed $r_H$
there exists a length $L$
 above which it becomes entropically favourable for
the mass to localize into a black hole with a $spherical$ horizon topology.
As such, some uniform strings decay into black holes, since the latter have higher
entropy. This connects, dynamically, phases 1 and 3 above, as shown by numerical evolutions~\cite{Lehner:2010pn}.
The non-uniform black strings, phase 2,
provide the connecting (non-dynamical) solution, as 
 they merge 
 with the branch of caged/localized black holes
  at a topology changing transition
\cite{Kudoh:2004hs,Kleihaus:2006ee}.

\medskip

The beautiful picture just described occurred entirely in vacuum for black objects. May it be replicated for some classes of self-gravitating horizonless objects, and hence non-vacuum?

It is well known that there is a resemblance between the GL instability and the Rayleigh-Plateu instability in fluids~\cite{Cardoso:2006ks}. 
Moreover, recently, it was shown that there is a  Rayleigh-Plateu type instability for $Q$-strings~\cite{Chen:2024axd}, which are Minkowski spacetime solitons in $D=(2+1)$-dimensions, augmented by a trivial flat direction \cite{Volkov:2002aj}. 
 This suggests a GL type instability, and a corresponding phase picture similar to the aformentioned one for black objects, also occurs in $D=(3+1)+1$ dimensions, in particular for the
 self-gravitating cousins of $Q$-balls, which are 
 the \textit{Boson Stars}~\cite{Schunck:2003kk}. Indeed, we shall show this is so in this paper.

 We consider a simple 
$D=5$ non-vacuum, non black object model: solitons in Einstein-Klein-Gordon (EKG) theory.
Considering again a spacetime
with standard KK asymptotics,
the simplest horizonless solutions of this 
model describe {\it uniform boson strings} (some times called \textit{vortices}). 
They can be viewed as natural KK generalizations
of the 
$D=4$  Boson Stars (BSs),  
which are finite energy, localized 
solutions of the EKG equations 
\cite{Kaup:1968zz,Ruffini:1969qy}. 
These horizonless configurations
provide a concrete realization of 
Wheeler's 'geon' idea \cite{Wheeler:1955zz},
being
usually regarded as 
'macroscopic quantum states' prevented from gravitationally collapsing by Heisenberg's uncertainty
principle.  
BSs have enjoyed constant attention over the last decades,
being used in a wide variety of models,
$e.g.$ as sources of dark
matter or as black hole mimickers
\cite{Liebling:2012fv}.

Before analysing the self gravitating boson strings, we shall dwell on a simpler model. Indeed, some of the basic BS features
are captured by simpler $D=4$ flat spacetime solutions: $Q$-balls
\cite{Coleman:1985ki}.
These are non-gravitating and  non-topological
solitons, obtained in a model with
a non-renormalizable self-interaction, arising in some effective field theories \cite{shnir}.
They circumvent the standard Derrick-type argument 
\cite{Derrick:1964ww}
by virtue of having a time-dependent phase for the scalar field. 
When viewed in the context of this work, 
the $D=4$  non-gravitating $Q$-balls 
can be trivially uplifted to
$D=5$, 
becoming  \textit{uniform  $Q$-strings}.
The uniform $Q-$strings are unstable to linearized perturbations with a long wavelength along the circle.
More precisely, for any configuration,
there is a critical size $L_0$  such that the strings
with $L \leq L_0$ are stable and those with $L>L_0$ are unstable. Observe the analogy with the GL case, the 
object's length scale being provided in this case
by the field's mass.

The latter possess a zero-mode 
with a non-trivial dependence of the extra-dimension  
for a critical value $L_0$.
This implies  
the branching
off into a
new family of {\it non-uniform $Q$-strings}  at the onset of the instability.

Restricting to  the simplest
static configurations, the results in this work
show the existence of three classes  of solutions, in (partial) analogy with the vacuum KK black strings and black holes discussed above.

\begin{itemize}
\item 
{\bf Uniform strings }
\\
These solutions have no $z-$dependence.
While the $D=4$ $Q$-balls trivially extend
into the extra-dimension,
the EKG case has been discussed in a more general context
in the recent work \cite{Brihaye:2023vox}.
As expected, they share most of the properties of the
four-dimensional boson stars.

\item 
{\bf Non-uniform strings }
\\
These are the non-linear continuation of the aforementioned zero-modes, emerging smoothly from the uniform solutions.
For a given length of the extra-dimension, they form a sequence which can be labeled in terms of the field frequency $\omega$.
Along this sequence, the non-uniformity of the solutions
increases, the configurations becoming
localized in the extra-dimension 
(although still with a single component, as in the uniform case).

\item 
{\bf Localized solitons }
\\
The uniform and non-uniform strings 
form the {\it fundamental} configurations.
In addition, there exist a class of 
 intrinsic five dimensional solutions, 
which form a disconnected  phase. 
Such configurations  possess an internal structure,
with $2n$ components localized on the $S^1$-circle (where $n\geq 1$),
and are
supported by 
excited KK scalar modes.
These modes provide a non-zero contribution
to the field's effective mass,
such that, different from  
the case of fundamental configurations, 
the localized solitons  exist as well for a  
massless scalar field.
  
\end{itemize}

\begin{figure}[h]
	\makebox[\linewidth][c]{%
		\begin{subfigure}[b]{8cm}
			\centering
\includegraphics[width=8cm]{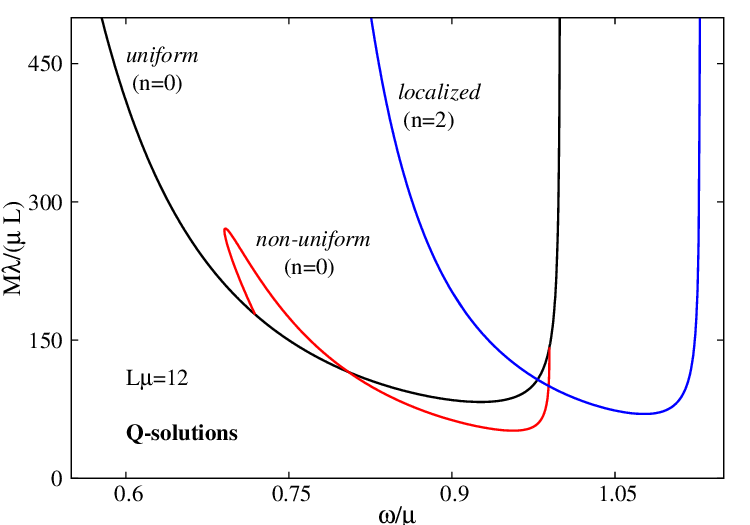}
		\end{subfigure}%
		\begin{subfigure}[b]{8cm}
			\centering
\includegraphics[width=8cm]{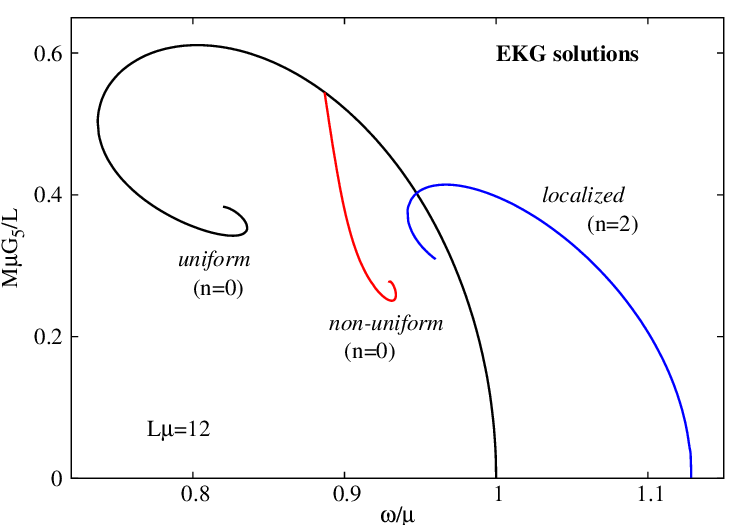}
		\end{subfigure}%
	}
 \\
	\caption{
	{\small
The generic phase diagram of solutions 
is shown for a model with a self-interacting scalar field 
(left panel) and a gravitating scalar field 
(right panel).}
		\label{phases}}
\end{figure}

The emerging picture for all these three phases of flat spacetime configurations is shown in Figure \ref{phases} (left panel). The diagram anticipates the results in Section \ref{toy}, namely that the uniform boson strings have a similar phase diagram as $Q$-balls and that the non-uniform strings branch off from their domain of existence, whereas the localized solutions have, again, a similar phase diagram to the uniform case. This qualitatively resembles the results in the self-gravitating case, which are obtained in Section \ref{EKG}, which we also anticipate in Figure \ref{phases} (right panel). Now the phase space of the uniform strings mimics that of BSs, with the other phases behaving in a way similar to that of the non-gravitating examples.

The existence of these different phases of the higher dimension Klein-Gordon and EKG system, raises the question of how to classify solutions. The number of nodes
of the scalar field (real) amplitude $\psi (r,z)$
 $w.r.t.$ the $z$-coordinate
(with $r$ the radial coordinate on ${\cal M}_4$)
provides a possible classification scheme.
Following \cite{nr},  
this index can be computed 
as\footnote{Note that the spacetime geometry
does not enter this definition.}
\begin{eqnarray} 
\label{n}
    n=\frac{1}{2 \pi} \int_{0}^{L}
    \frac{\psi_{,z}(r_0,z)^2-\psi(r_0,z)\psi_{,zz}(r_0,z)}
    {\psi_{,z}(r_0,z)^2+\psi(r_0,z)^2}
        dz~,
\end{eqnarray}
with $r_0$ arbitrary (in particular $r_0=0$).
 The  strings  have $n=0$
 (no $z-$nodes),
 while the index for
 localized solitons is $2n$.

\medskip

The outline of the paper is as follows. 
In Section \ref{toy},
we consider a simple toy model  
with a complex scalar field with selfinteraction 
in a fixed  ${\cal M}_4\times S^1$
background.
The  solutions there captures most of the basic properties
of those of the EKG system
(with a massive, complex scalar field without selfinteraction),
which are discussed in Section
\ref{EKG}. 
We conclude with Section \ref{final}, where the results
are compiled, together with possible avenues for future
research.  

\medskip
 {\bf Conventions.}
%
Throughout the paper, early Latin letters, $a,b,\dots $ 
label five dimensional spacetime coordinates,
running from $0$ to $4$ (with $x^0=t$ and $x^4=z$ the extra-dimension),
  while middle letters, $i,j,\dots $ 
are used for $D=4$ spacetime coordinates, running from $0$ to $3$.
Also, as standard, we use Einstein's summation convention.

The background of the model
corresponds to four dimensional Minkowski spacetime times a circle,
 ${\cal M}_4\times S^1$, 
with a line element 
\begin{eqnarray}
ds^2=-dt^2+dr^2+r^2 d\Omega_2^2+dz^2,
~~{\rm with}~~d\Omega_2^2=d\theta^2+\sin^2 \theta d\varphi^2~,
\end{eqnarray}
$(r, \theta,\varphi)$ 
being the usual spherical coordinate, 
$t$ is the time and  $z$ is
a which is compactified on a circle,
 $z\sim z + L$.

\section{$Q$-strings on a fixed 
${\cal M}_4 \times S^1$ background }
\label{toy}

\subsection{The $D=5$ model}
 
We consider the usual action for
a complex scalar field $\Psi$, in $D=5$,
\begin{eqnarray}
\label{actionQ}
S= \int  d^5x \sqrt{-g}\left[ 
   \frac{1}{2} g^{ab}
   \left( \Psi_{, \, a}^* \Psi_{, \, b} + \Psi _
{, \, b}^* \Psi _{, \, a} \right) 
+U(|\Psi| ) 
 \right] ,
\end{eqnarray}
where $U(|\Psi| )$ is the scalar field potential,
with the Klein-Gordon equation
\begin{eqnarray}
\label{KG}
\nabla_{a} \nabla^{a}\Psi=\frac{d U}{d\left|\Psi\right|^2} \Psi ,
\end{eqnarray}
and the following expression of the
 energy-momentum tensor 
\begin{eqnarray}
\label{tmunu} 
T_{ab} 
=
\partial_a\Psi^*\partial_b\Psi 
+\partial_b\Psi^* \partial_a\Psi  
-g_{ab}  \left[ \frac{1}{2} g^{cd} 
 ( \partial_c \Psi^* \partial_d\Psi+
 \partial_c\Psi^*  \partial_d\Psi) +U(\left|\Psi\right|) \right]
 \ .
\end{eqnarray} 
In what follows we shall focus on the  
simplest  scalar
potential in the $Q$-ball literature~
\cite{Coleman:1985ki}:
\begin{eqnarray}
\label{potential}
U(\Psi)=\mu^2|\Psi|^2-\lambda |\Psi|^4+\nu |\Psi|^6 ~,
\end{eqnarray}
where
$\mu$ is the scalar field mass
and
$\lambda,\nu$ are positive parameters 
controlling the self-interactions of the scalar field.
The ground state of the model is $\Psi=0$, 
being approached as $r\to \infty$.

We shall (mainly) consider
a scalar field ansatz with
 \begin{eqnarray}
 \label{psin}
 \Psi \equiv  \psi(r,z) e^{-i\omega t},~~
 {\rm where}~~\psi(r,z)=\psi(r,z+L),
\end{eqnarray}
with $\omega>0$
  the frequency
and $\psi$
 the amplitude,
which is  $real$. 
From (\ref{KG}), $\psi$ solves the equation  
\begin{eqnarray}
\label{psieq}
 \psi_{,rr}+\frac{2\psi_{,r}}{r}+ \psi_{,zz}
(\omega^2-\mu^2)\psi+2\lambda \psi^3-3\nu \psi^5=0~.
\end{eqnarray}

The energy density of a configuration reads
\begin{eqnarray}
\rho \equiv -T_t^t = \psi_{,r}^2+\psi_{,z}^2+(\omega^2+\mu^2)\psi^2-\lambda \psi^4
+\nu \psi^6~,
\end{eqnarray}
with the total mass-energy 
\begin{eqnarray}
M= 4 \pi   \int_0^L dz \int_{ 0}^\infty dr ~ r^2 \rho \ .
\end{eqnarray}

Let us remark that
the action (\ref{actionQ}) is invariant under the {\it global} $U(1)$ transformation
$\Psi\rightarrow e^{i\alpha}\Psi$, where $\alpha$ is a constant.
This implies that the current 4-vector 
$j^a=-i (\Psi^* \partial^a \Psi-\Psi \partial^a \Psi^*)$
is conserved, $i.e.$ $j^a_{\ ;a}=0$.
Therefore, integrating the timelike component of this current on a spacelike slice $\Sigma$ yields a conserved quantity -- the \textit{Noether charge}:
\begin{eqnarray}
	\label{Q}
	Q_N=\int_{\Sigma}~j^t = 8 \pi \omega \int_0^L dz \int_{ 0}^\infty dr ~r^2 \psi^2(r,z).
\end{eqnarray} 

The $\psi$-equation (\ref{psieq})
can also be obtain from the following  effective action 
\begin{eqnarray}
\label{Seff}
S(\psi)=\int_0^\infty dr~r^2
\int_0^L dz
\left[
\psi_{,r}^2+\psi_{,z}^2+U(\psi)-\omega^2\psi^2
\right],
\end{eqnarray}
which allows to derive a  virial relation 
by using a Derrick-type scaling technique 
\cite{Derrick:1964ww}, as follows.
Let us assume the existence of a solution $\psi(r)$
 with suitable boundary conditions at the origin and at infinity. 
 Then each member of the 1-parameter family
 $\psi_{\lambda} (r,z) \equiv \psi (\lambda r,z)$ 
assumes the same boundary values at $r=0$ and $r=\infty$, and the action 
$S_{\lambda} \equiv S[ \psi_{\lambda}]$
must have a critical point at $\lambda=1$, $[dS/d\lambda]_{\lambda=1}=0$.
Thus  we obtain the virial relation 
\begin{eqnarray}
\label{virial}
 \int_0^\infty dr~r^2
\int_0^L dz
\left[
\psi_{,r}^2+3(\psi_{,z}^2+U(\psi)-\omega^2\psi^2)
\right]=0.
\end{eqnarray}
One notices that the $z$-dependence plays the same role as the potential, the solutions being always supported 
by the harmonic  time dependence of the
scalar field.

\subsubsection{The $D=4$ effective model}

The $D=5$ non-uniform solutions with the scalar field ansatz
(\ref{psin})
can also be considered from a $D=4$ perspective.
The corresponding 
$D=4$ 
description is found by taking the 
Fourier decomposition of the scalar amplitude
\begin{eqnarray}
\label{Fourier}
 \psi(r,z)= \sum_{m \geq 0}  \psi_m(r)\cos (m kz)~,
~~~~{\rm with}~~~k=\frac{2\pi }{L},
\end{eqnarray}
where 
$\phi_m(r)$ is the amplitude of a given mode $m$.
After replacing in the model's action
(\ref{actionQ})
and integrating over the $z$-direction,  
it follows that the five-domensional 
solutions
can be interpreted 
as describing $Q$-ball-type
solitons in a $D=4$ model 
(in Minkowski spacetime background),
with an infinite set of interacting massive 
scalars, all of them with the same harmonic time-dependence.
The corresponding $D=4$ Lagrangian density reads
 \begin{eqnarray}
{\cal L}={\cal L}_0
+\frac{1}{2}\sum_{m\geq 1} {\cal L}_{m}+
{\cal L}_{int}~,
\end{eqnarray}
with
 \begin{eqnarray}
 &&
 {\cal L}_0=\psi^2_{0,r}+(\mu^2-\omega^2)\psi_0^2
 -\lambda \psi_0^4+\nu \psi_0^6,~~
 \\
 \nonumber
 &&
{\cal  L}_{m}=  
 \psi^2_{m,r}+(\mu^2+m^2 k^2-\omega^2)\psi_m^2
 -\frac{3}{4}\lambda \psi_m^4+\frac{5}{8}\nu \psi_m^6,~~ 
 \\
 \nonumber
 &&
{\cal  L}_{int}=
 \lambda \sum_{i_s \geq 0} c^{(4)}_{i_1, i_2, i_3, i_4} 
 \psi_{i_1}  \psi_{i_2} \psi_{i_3} \psi_{i_4}
 + 
 \nu \sum_{i_s \geq 0} c^{(6)}_{i_1,i_2, i_3, i_4, i_5, i_6} 
 \psi_{i_1}  \psi_{i_2} \psi_{i_3} \psi_{i_4} \psi_{i_5} \psi_{i_6}~,
\end{eqnarray}
 where no general expression appears to exist for
 the coefficients
 $c^{(4)}_{i_1, i_2, i_3, i_4} $,
 $c^{(6)}_{i_1, i_2, i_3, i_4, i_5, i_6} $
 (with
 $e.g.$
 $c^{(4)}_{0,0,1,1}=-3$
 and
  $c^{(6)}_{0,0,0,0,1,1}=15/3$).
 Note that each field
 possesses an effective mass
 $\mu_m^2=\mu^2 +m^2 k^2$,
 with an extra-KK contribution.

\subsubsection{Scaling and units}

Inspection of the  equation  (\ref{psieq}), shows the existence of the following scaling
symmetries of the  model
\begin{eqnarray}
 &&
i)~~ \psi \to a \psi,~~\lambda \to \lambda/a^2,~~
\nu \to \nu/a^4,
\\
&& 
ii)~~ r\to a r,~~z\to a z,~~ 
\mu  \to  \mu/a ,~~
\lambda \to  \lambda/a^2 ,~~
\nu \to   \nu/a^2,~~
\omega \to  \omega/a ,~~
\end{eqnarray}
where $a$ is an arbitrary non-zero parameter.
These symmetries are used  
 to set $\mu=1$, $\lambda=1$
 in the numerics,
 $i.e.$ one takes
\begin{eqnarray}
r\to r /\mu\ ,  \qquad 
z\to z /\mu\ ~~{\rm and}~~
\phi \to \phi \mu/\sqrt{\lambda}~,   
\end{eqnarray} 
together with $L\to L /\mu$, $\nu \to \mu\lambda^2/\mu^2$.
As such,  the only input parameters are 
$i)$  the field frequency $\omega$,
$ii)$   the (rescaled) coefficient  $\nu$ of the sextic term in the
scalar potential, and
$iii)$  the size $L$ of the extra-dimension 
(which, however, does not enter the equations).
All results reported in this Section are found for 
the choice
$\nu=0.275$ (in which case the scalar potential $U$ is strictly positive).
Also, all functions and quantities of interest are expressed in units set by $\mu$ and $\lambda$.

\subsection{Uniform $Q$-strings: zero mode and  stability}
\label{zm}

\subsubsection{The zero mode }
\label{zero}

All known $D=4$ 
$Q$-ball solutions 
can be promoted to solutions of the five-dimensional model (\ref{actionQ}), 
becoming
\textit{uniform
$Q$-strings}.
 The simplest such solutions are
 spherically symmetric (from a $D=4$ perspective), 
 with
 $ \psi \equiv  \psi(r )$ 
 in (\ref{psin}).
Their properties have been extensively studied in the literature  \cite{shnir},
despite the absence of exact solutions.
It follows that $\omega$ necessarily belongs to the interval
$\omega_{min}<\omega <\mu$. 
Both the mass and the Noether charge diverge at the
limits of this interval, the families of solutions being disconnected from the $\psi=0$ ground state -
see the corresponding curves in Figures 
\ref{phases}, \ref{wM}.
Also,
given a $\omega$-value,
the $Q$-balls exist for a discrete set of the
scalar field at the origin, $\psi(0)$,
which are indexed by the number of nodes of the scalar amplitude.
In what follows we shall restrict to
the case of fundamental, nodeless solutions.

Apart from these, there are 
also
$D=5$  solutions with a
 dependence of the extra-dimension -- the {\it non-uniform Q-strings}.
They can be taken as non-linear 
continuation of the zero-modes around
the uniform configurations.
To study the zero modes, one considers
 a scalar field ansatz with
 \begin{eqnarray}
\label{pert1}
\psi(r,z)=  \phi_0(r)+\epsilon \phi_1(r) \cos (m k z),  
\end{eqnarray}
where
$\phi_0(r)$ describes the uniform solution,
$\epsilon$
a small parameter
and
  $m=\pm 1, \pm 2, \dots$.
	In what follows, we shall restrict to the $m=1$ case, since
	the solutions with higher $m$ can be generated by taking $L\to L/m$, which is a symmetry of the problem.  

The linearized equation (\ref{psieq})
implies that
the functions $\phi_0(r)$ and 
 $\phi_1(r)$
solve the equations
\begin{eqnarray}
&&
\phi_0''+\frac{2\phi_0'}{r}+(\omega^2-\mu^2)\phi_0+2\lambda \phi_0^3-3\nu \phi_0^5=0
\\
&&
\label{eq1}
\phi_1''+\frac{2\phi_1'}{r}+(\omega^2-\mu^2-k^2)\phi_1
+(6\lambda \phi_0^2-15\nu \phi_0^4)\phi_1=0,
\end{eqnarray}
with the following approximate small-$r$ expression
\begin{eqnarray} 
\label{as1}
&&
\phi_0(r)=\phi_0(0)+\frac{1}{6}\phi_0(0)
\left(
\mu^2-\omega^2-2\lambda \phi_0(0)^2+3\nu \phi_0(0)^4
\right) r^2+\dots,
\\
&&
\phi_1(r)=\phi_1(0)+\frac{1}{6}\phi_1(0)
\left(
k^2+\mu^2-\omega^2-6\lambda \phi_0(0)^2+15 \nu \phi_0(0)^4
\right) r^2+\dots,
\end{eqnarray}
while for large-$r$ one finds
\begin{eqnarray} 
\label{as2}
\phi_0(r)= c_0\frac{e^{-\sqrt{\mu^2 -\omega^2}r}}{r}+\dots
~~
\phi_1(r)= c_1\frac{e^{-\sqrt{\mu^2 +k^2-\omega^2}r}}{r}+\dots
\end{eqnarray}
where 
$\phi_0(0)$, $c_0$, $c_1$
are parameters fixed by numerics;
also, since the equation for $\phi_1$
is linear,
we set $\phi_1(0)=1$
without any loss of generality.

Therefore the problem consist in solving an eigenvalue problem for the
function  
$\phi_1(r)$
in a background given by a uniform $Q$-string with a given $\omega$.
Restricting to the fundamental states
($i.e.$ no $r$-nodes also for $\phi_1$),
the solution smoothly interpolating between the above asymptotics
exists for a single value of $k=2\pi/L$, 
which is found by using a standard shooting procedure.

\begin{figure}[h!]
\begin{center}
\includegraphics[width=0.7\textwidth]{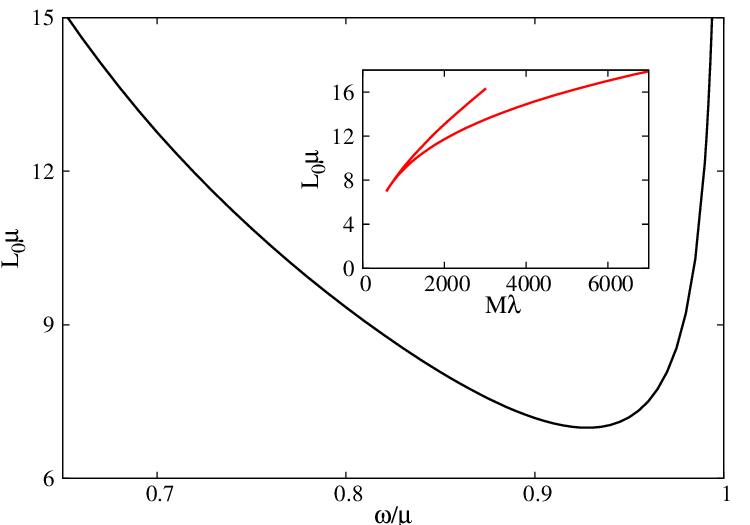} 
\caption{ 
The critical length $L_0$ of the
$z$-direction allowing for 
a scalar field zero-mode is shown as a function of the field frequency
$\omega$ (main panel),
and of the total mass $M$ (inset), of the
uniform $Q$-strings. 
}
\label{mode}
\end{center}
\end{figure}

The numerical results are shown in Figure \ref{mode}.
As one can see,
for any uniform $Q$-string (as specified $e.g.$ by
the field frequency
$\omega$), 
there exist a unique critical value of the length 
$L \equiv L_0$ of the extra-dimension
which allows for a zero mode.
Interestingly, one finds an overall minimal value of $L_0$, with
\begin{eqnarray}
\label{bound}
L_0>L_{min} \simeq \frac{6.989}{\mu},
\end{eqnarray}
which occurs for
$\omega/\mu \simeq 0.93$.
Moreover, the value of $L_0$
appears to increase without bound at the limits
of the $\omega$-interval.
Also, for a given value of $L>L_{min}$,
there are two different uniform $Q$-strings
possessing a zero mode with the same wavenumber $k$.
Alternatively, a solution with a given mass 
possesses $two$ zero modes.

\subsubsection{The linear stability of the uniform $Q$-strings}

As $D=4$ solutions,
a subset of $Q$-balls are known to be stable 
\cite{Coleman:1985ki,PaccettiCorreia:2001wtt,Sakai:2007ft}. Promoted to $D=5$ $Q$-strings, 
however, those very solutions (classified, say by their frequency) could become unstable $w.r.t.$
perturbations depending on the extra-dimension.
Also, the existence of a zero mode 
(discussed above)
is
usually associated to
the threshold between stable and unstable configurations.

The issue of linear stability of the uniform $Q$-strings  can be approached as follows 
\cite{Panin:2016ooo}.
Instead of 
(\ref{pert1}),
(\ref{psin}), 
we start with a more general scalar field Ansatz 
with a {\it complex} amplitude,
\begin{eqnarray}
\label{p1}
\Psi =  (\phi_0(r)+\epsilon ( u(r)+i v(r)) \cos ( k z) e^{\Omega t} ) e^{-i \omega t}.
\end{eqnarray}
with 
 $\Omega$ a  parameter. 
The (real) functions
$u$ and
$v$
solve the equations
\begin{eqnarray}
&&
\label{eqs1}
 u''+\frac{2u'}{r}+(\omega^2-k^2-\Omega^2)u
-(\mu^2-6 \lambda \phi_0^2+15 \nu \phi_0^4)u
-2\omega\Omega v=0,
\\
&&
\label{eqs2}
v''+\frac{2v'}{r}+(\omega^2-k^2-\Omega^2)v
-(\mu^2-2 \lambda \phi_0^2+3 \nu \phi_0^4)v
+2\omega\Omega u=0.
\end{eqnarray}
Note that the zero-mode solutions in Section \ref{zm} are
recovered for $v=0$, $\Omega=0$
and $u\equiv \phi_1$.

Given a $Q$-ball solution with some
$\omega$,
the equations 
(\ref{eqs1}), (\ref{eqs2}) 
are solved with the boundary conditions
\begin{eqnarray}
u(0)=1,~v(0)=v_0,~~u(\infty)=v(\infty)=0,
\end{eqnarray}
(with $v_0$ a parameter fixed by numerics),
which results in an eigenvalue problem for $(\Omega,k)$.

\begin{figure}[h!]
\begin{center}
\includegraphics[width=0.7\textwidth]{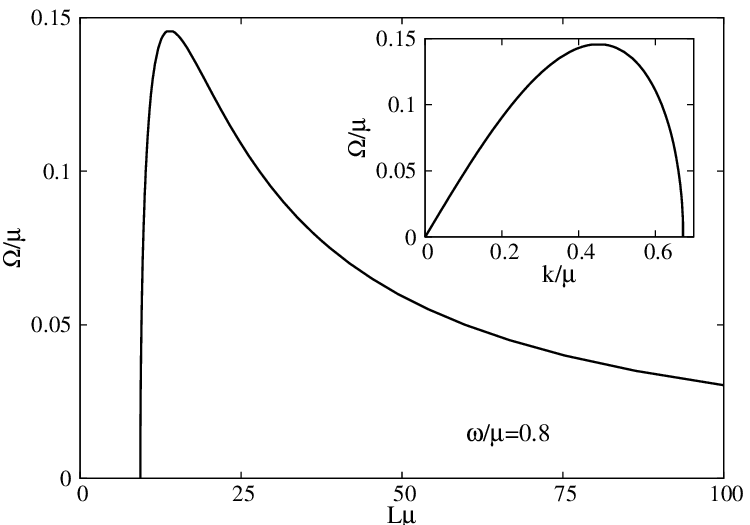} 
\caption{ 
Eigenvalues 
$(L,\Omega)$ 
and
$(k,\Omega)$ (inset) 
for which an
instability is present,  
for a typical
uniform $Q$-string solution. 
 }
 \label{Omega}
\end{center}
\end{figure}

The numerical results for a background
$Q$-string with $\omega/\mu=0.8$
are shown in Figure \ref{Omega}.
Similar results were found for few other values of $\omega$, such that 
 picture in Figure \ref{Omega} 
 is likely to be generic.
 This confirm the expectation that,
for a given $\omega$,
 all solutions with $L>L_0(\omega)$ -
with $L_0(\omega)$ the value for the zero mode in Figure~\ref{mode} - 
are unstable, $i.e.$ $\Omega^2>0$.
As such, no instability is found
for  configurations with
a small enough size of the extra dimension
$L_0<L_{min}$
as given by
(\ref{bound}). 
This is in agreement with the recent results in~\cite{Chen:2024axd}.
 
\begin{figure}[h]
	\makebox[\linewidth][c]{%
		\begin{subfigure}[b]{8cm}
			\centering
\includegraphics[width=8cm]{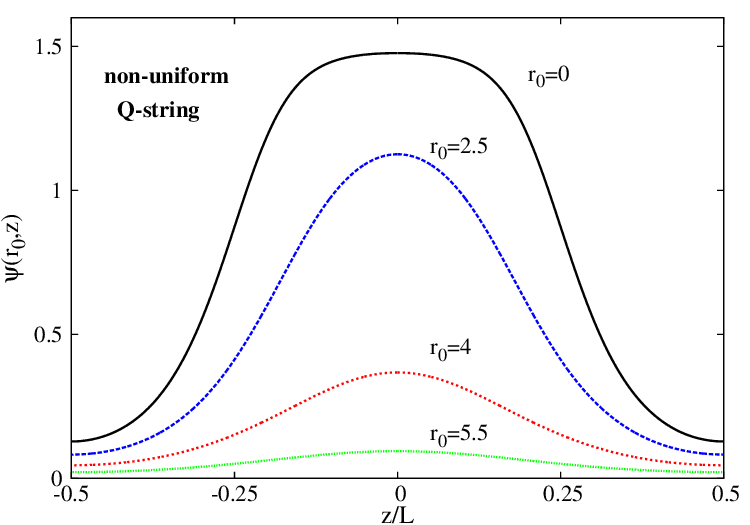}
		\end{subfigure}%
		\begin{subfigure}[b]{8cm}
			\centering
\includegraphics[width=8cm]{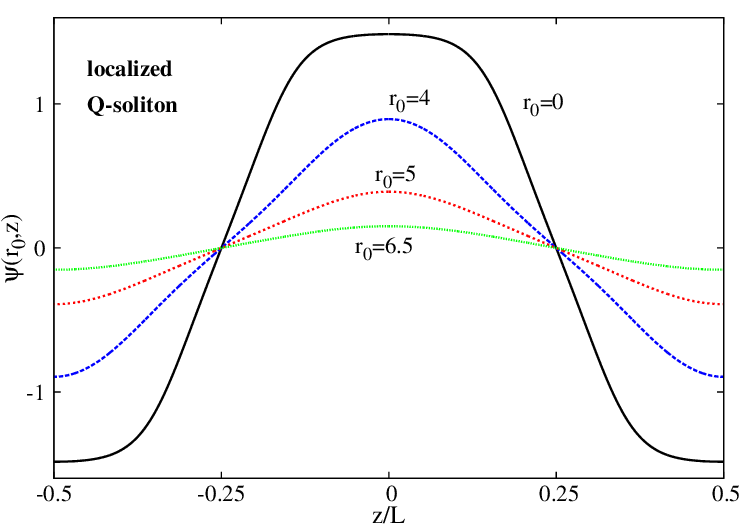}
		\end{subfigure}%
  } 
  \\
  \\
 \makebox[\linewidth][c]{%
		\begin{subfigure}[b]{8cm}
			\centering
\includegraphics[width=8cm]{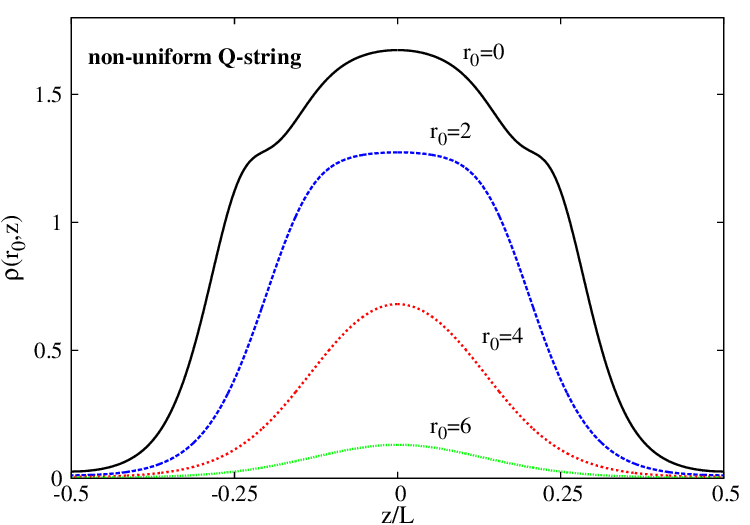}
		\end{subfigure}%
		\begin{subfigure}[b]{8cm}
			\centering
\includegraphics[width=8cm]{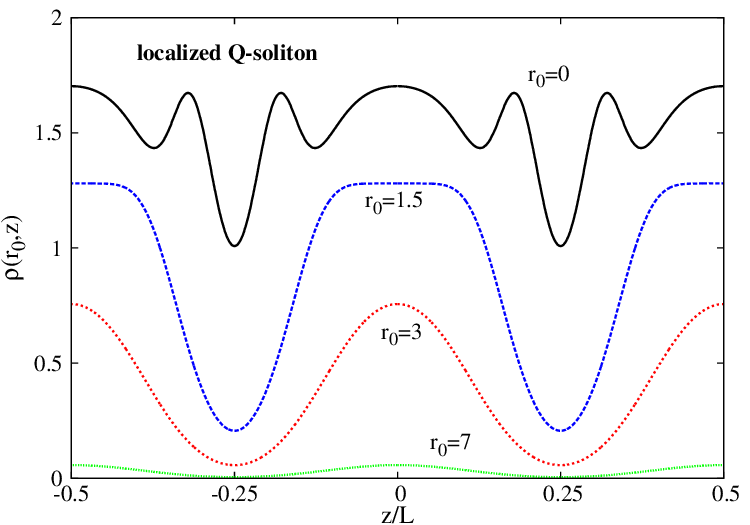}
		\end{subfigure}%
	}
 \\
	\caption{
	{\small
	The scalar field amplitude
 $\psi$  
 and the energy density $\rho=-T_t^t$
 of a typical non-uniform $Q$-string
 as a function of the 
$z$-coordinate for several fixed values of $r$.
 Here and in Figure  
 \ref{Zfixed-z-Q} 
 one considers two solutions with
 the same input parameters
 $L \mu=12$,
 $\omega/\mu=0.7$.  
}
		\label{Zfixed-r-Q}}
\end{figure}
%
\subsection{The non-uniform $Q$-strings}

The non-linear continuation of the zero-modes 
discussed in Section \ref{zero}
results
in {\it non-uniform} $Q$-strings.
They are found by solving numerically
the equation (\ref{psieq}) 
with the following  boundary conditions
\begin{eqnarray}
\partial_r \psi \big |_{r=0}=\psi \big |_{r=\infty}=0,~
\partial_z \psi \big |_{z=0,L}=0.
\end{eqnarray}

The numerical approach is similar to that 
used in the EKG case (below), while, as expected,
the  accuracy is much better for these flat spacetime solutions.

\begin{figure}[h]
	\makebox[\linewidth][c]{%
		\begin{subfigure}[b]{8cm}
			\centering
\includegraphics[width=8cm]{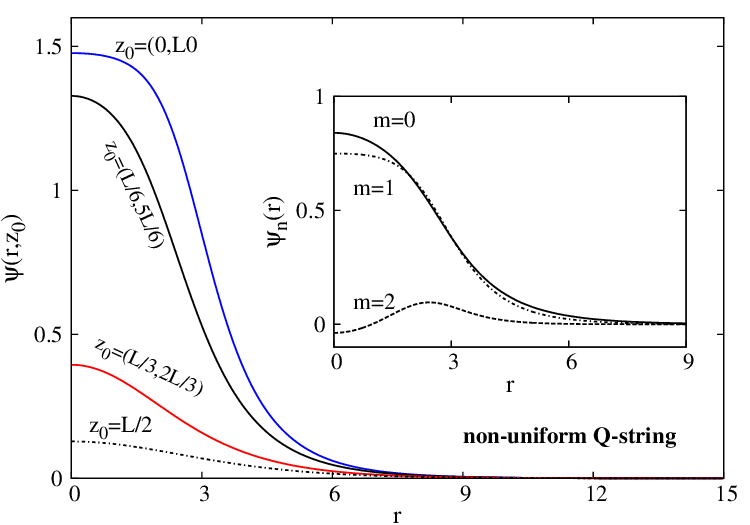}
		\end{subfigure}%
		\begin{subfigure}[b]{8cm}
			\centering
\includegraphics[width=8cm]{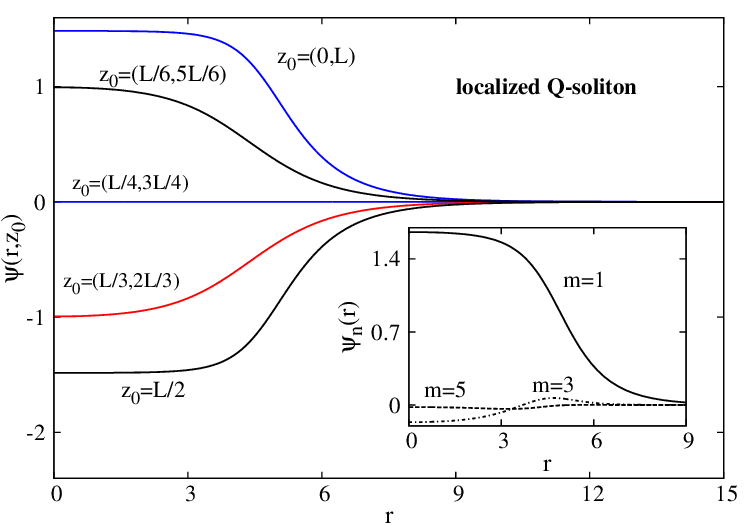}
		\end{subfigure}%
   } 
  \\
  \\
 \makebox[\linewidth][c]{%
		\begin{subfigure}[b]{8cm}
			\centering
\includegraphics[width=8cm]{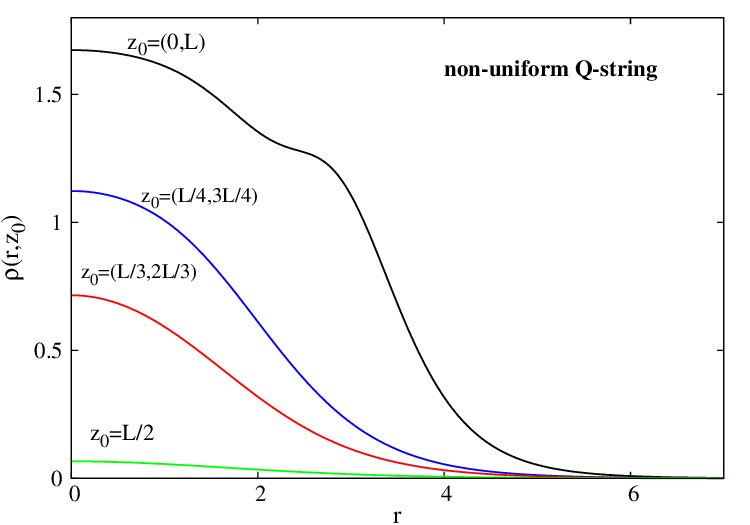}
		\end{subfigure}%
		\begin{subfigure}[b]{8cm}
			\centering
\includegraphics[width=8cm]{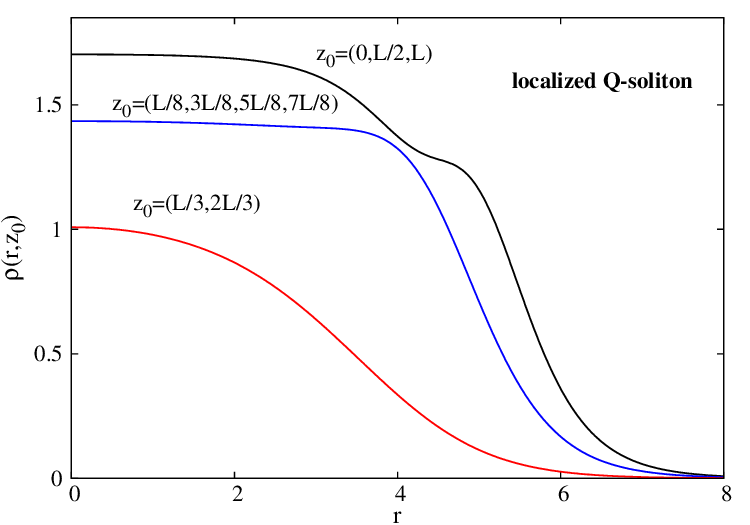}
		\end{subfigure}%
	}
 \\
	\caption{
	{\small
The scalar field amplitude
 $\psi$  
 an the energy density   $\rho=-T_t^t$ 
  of a typical non-uniform $Q$-string
 as a function of the  
 radial
 coordinate for several fixed values of the extra-dimension $z$.
The insets show the mode decomposition
of the scalar field amplitude according to eq.
(\ref{Fourier}).  
}
		\label{Zfixed-z-Q}
  }
\end{figure}

As expected, all zero-modes
possess non-linear continuations, which therefore 
exist  for value of $L$ above the bound (\ref{bound}),
only.
The profile of 
the scalar field amplitude
and of the energy density
of
a typical non-uniform solution are shown in Figures 
\ref{Zfixed-r-Q} and  \ref{Zfixed-z-Q} 
(left panels),
for fixed values of $r$
and $z$, respectively. 

While the $r-$dependence of solutions is rather similar
to that found in the uniform case, they are localized 
on the $z-$circle, with a single component at $z=0$.

Also, let us remark that, when considering
a Fourier decomposition of the field amplitude, 
eq. (\ref{Fourier}), 
all modes are present (although $m=1$
provides the dominant contribution, see the inset
in Figure \ref{Zfixed-z-Q}).

\begin{figure}[h!] 
	\makebox[\linewidth][c]{%
		\begin{subfigure}[b]{8cm}
			\centering
\includegraphics[width=8cm]{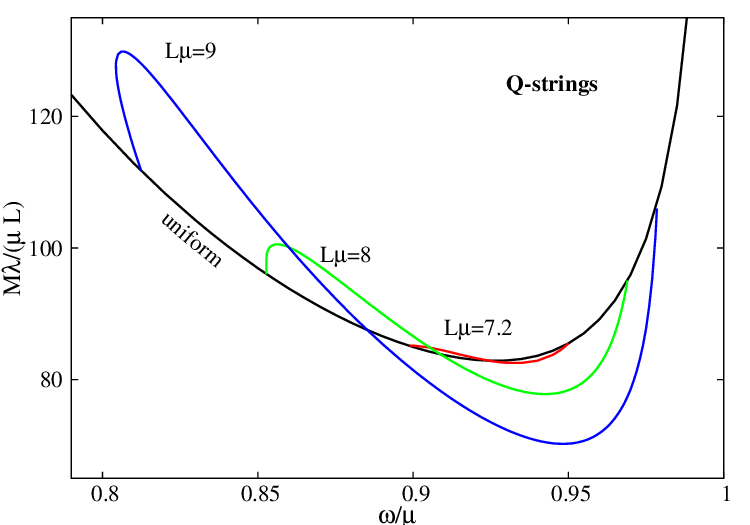}
		\end{subfigure}%
		\begin{subfigure}[b]{8cm}
			\centering
\includegraphics[width=8cm]{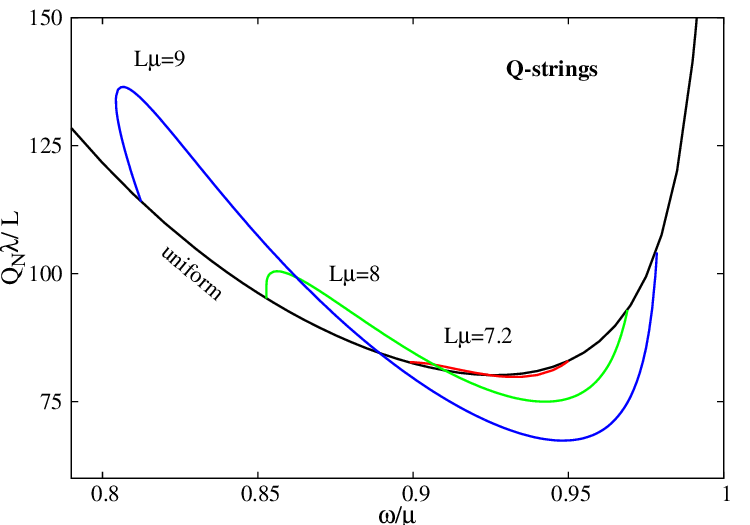}
		\end{subfigure}%
	}
\caption{ 
The mass $M$ and Noether charge $Q_N$ 
of $Q$-strings are shown as a function of the field frequency
$\omega$.
The non-uniform branches bifurcate from the uniform set (black curve).
Results for
 three different values of the parameter $L$
are displayed.  
}
\label{wM} 
\end{figure}

As one can anticipate from Figure \ref{mode},
in a mass-frequency diagram, the
non-uniform $Q$-string interpolate between
two different uniform solutions, see Figures 
\ref{phases}, \ref{wM}.

\subsection{The localized $Q$-solitons}

These configurations are intrinsic five dimensional
and are $not$ connected to the strings\footnote{Strictly speaking, 
the non-uniform strings are also localized configurations.
However, since their
non-uniformity can be arbitrarily small, 
we reserve this terminology 
for the solutions in this Subsection.}.
The starting point here is to remark that,
for a given $L$, the linearized Klein-Gordon
equation possess the (single mode) solution
 (with $ m\geq 1$)
\begin{eqnarray}
\label{sup1}
 \psi(r,z)= c_0 \cos(m kz) \frac{e^{-\sqrt{\mu^2+m^2 k^2-\omega^2}r}}{r},
\end{eqnarray}
with $c_0$ an arbitrary constant.
The localization condition
implies
an upper bound on frequency,
\begin{eqnarray}
\omega \leq \mu_{eff},~~{\rm with}~~
\mu_{eff}=\sqrt{\mu^2+n^2k^2 }~,
\end{eqnarray}
with a contribution from the $z$-dependence of the field.
As such,   configurations with
$\omega>\mu$ are allowed in this case.

The amplitude (\ref{sup1})
diverges at the origin, $r = 0$; 
however, as with the $m=0$ case, the  
non-linearities introduced by the sextic
scalar potential (\ref{potential})
regularize  this singularity,
(\ref{sup1}) being approached in the far field, only - observe the analogy with the $D=4$ 
multipolar boson stars
\cite{Herdeiro:2020kvf}.

The profile of a typical 
 solution is shown in Figures 
\ref{Zfixed-r-Q} and  \ref{Zfixed-z-Q}  
(right panels).
One can see that while the $r$-dependence 
of the scalar profile is strictly monotonic, $\psi$ possesses two nodes $(n=2)$ along the $z$-direction
(such that it can be taken as a
non-linear realization
 of the mode (\ref{sup1}) with $m=1$).
Also, this implies the
existence of two distinct components,
 localized $oppositely$ on the $z-$circle.
 
\begin{figure}[h!] 
	\makebox[\linewidth][c]{%
		\begin{subfigure}[b]{8cm}
			\centering
\includegraphics[width=8cm]{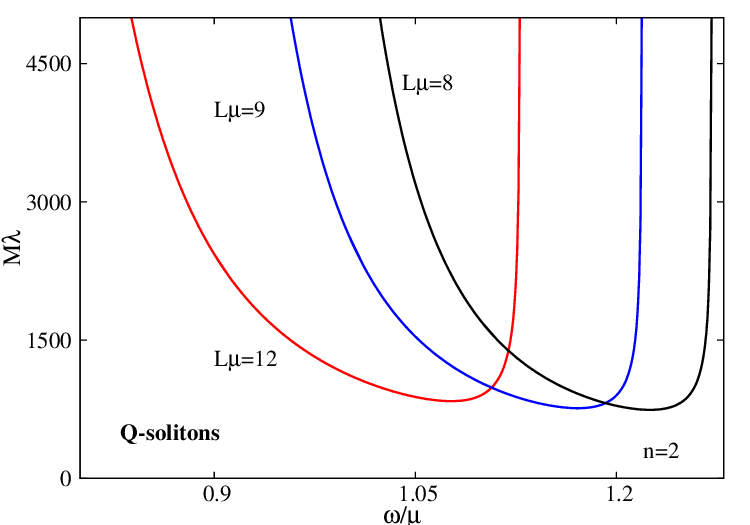}
		\end{subfigure}%
		\begin{subfigure}[b]{8cm}
			\centering
\includegraphics[width=8cm]{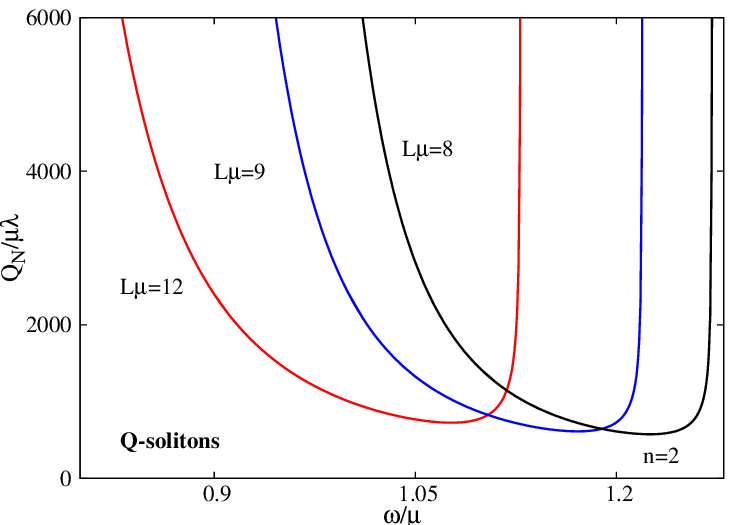}
		\end{subfigure}%
	}
\caption{ 
The (mass/Noether charge-frequency)-diagram  localized $Q$-solitons
with $n=2$ and three different values of the length $L$
of the extra-dimension. 
}
\label{wM-BS} 
\end{figure}

 The emerging picture  is summarized in Figure \ref{wM-BS}.
The solutions there have $n=2$; however, we have also found configuration 
with $n=4$,
which seem to follow a similar pattern. 
Note the strong similarity of the $(\omega,M)$-diagram with that found for uniform configurations;
for example,
the solutions exist for $\omega_{min}<\omega<\mu_{eff}$,
the mass and Noether charge diverging at the ends of this interval\footnote{In principle,
 the branches of non-uniform strings
 can also approach 
the frequency range $\omega>\mu$.
However, this does not seem too be
the case, the strings always
possessing a $m=0$-component in (\ref{Fourier}) inherited
from  the uniform configurations.}.
Also, one notices the existence of mass (and Noether charge) gap with respect to the $\psi=0$ ground state,
which reflects the non-perturbative nature of the $Q$-solitons.

%

Localized solitons
are likely to exist for any value of $L$ and 
all families of solutions studied so far are disconnected from the fundamental set of
$Q$-strings.

Finally, we mention the existence of localized solitons  with a massless scalar field, $\mu=0$.
The confining mechanism necessary is 
provided  by field's dependence on the
extra-dimension,
with 
  $\omega<\mu_{eff} = \frac{2\pi}{L}$. 
 The (mass/Noether charge-frequency)-diagram in this case is similar to  that displayed in Figure \ref{wM-BS}.

\section{The Einstein--Klein-Gordon solutions }
\label{EKG}
 
\subsection{The model}
 
We now turn on gravity and consider $D=5$ Einstein's gravity minimally coupled to a massive complex scalar field $\Psi$ \textit{without self-interaction}. Indeed, the coupling to gravity, and the non-linearities of the latter, are enough to allow the self-gravitating cousins of $Q$-balls, $i.e.$ BSs.
The system is described by the action  
\begin{equation}
\label{action}
S=\frac{1}{4 \pi G_5 }\int  d^5x \sqrt{-g}\left[ \frac{1}{4 }R
   -\frac{1}{2} g^{ab}\left( \Psi_{, \, a}^* \Psi_{, \, b} + \Psi _
{, \, b}^* \Psi _{, \, a} \right) 
- \mu^2 \Psi^*\Psi
%
 \right] ,
\end{equation}
where $G_5$ is Newton's constant.
 %
Variation of this action with respect to the metric and scalar
field gives the EKG equations:
\begin{equation}
\label{EKG-eqs}
 R_{ab}-\frac{1}{2}g_{ab}R=2 ~T_{ab},
~~\left(\nabla^2 -\mu^2\right)\Psi=0~,
\end{equation}  
where 
\begin{equation}
\label{Tab}
T_{ab}= 
 \Psi_{ , a}^*\Psi_{,b}
+\Psi_{,b}^*\Psi_{,a} 
-g_{ab}  \left[ \frac{1}{2} g^{cd} 
 \left( \Psi_{,c}^*\Psi_{,d}+
\Psi_{,d}^*\Psi_{,c} )+\mu^2|\Psi|^2
\right)
\right]~,
\end{equation}
is the
 stress-energy tensor  of the scalar field.

$D=5$ solutions of~\eqref{action}, 
with no-dependence on the extra-dimension, can be KK reduced to a $D=4$ configuration, using the KK metric ansatz:
\begin{eqnarray}
	\label{metric5}
 \nonumber
ds_5^2 = e^{-\alpha \phi (x)} ds_4^2+e^{2\alpha \phi (x)} (dz+2A_i(x) dx^i)^2,~
 {\rm with}~ds_4^2=g_{ij}^{(4)}(x) dx^i dx^j~~{\rm and}~~\alpha=\frac{2}{\sqrt{3}}.{~}
\end{eqnarray}
After integrating over the $z-$coordinate and
dropping  a boundary term,
the resulting $D=4$  action reads
\begin{eqnarray}
	\label{action4} 
	&&
	\mathcal{S}_4 = \frac{1}{4 \pi G_4}\int_\mathcal{M}  d^4x \sqrt{-g^{(4)}}\bigg[  
	\frac{1}{4}R^{(4)} 
	-\frac{1}{4} e^{3 \alpha \psi} F_{ij} F^{ij} 
	-\frac{1}{2} \partial_i\phi \partial^i \phi
	\\
	\nonumber
	&&
	{~~~~~~~~~~~~~~~~~}
	-\frac{1}{2} g^{ij(4)}
	\left(
	\partial_{i}\Psi^*  \partial_{j}\Psi +  
	\partial_{j}\Psi^*  \partial_{i}\Psi 
	\right) 
	-\mu^2 |\Psi|^2 e^{-\alpha \phi}
	\bigg] ~,
\end{eqnarray}
with
$G_4=G_5/L$, $\phi$ the dilaton and $F=dA$
a $U(1)$ field.
Note that while $F=0$ is a consistent truncation,
one can set $\phi=0$ in the vacuum case, only. 

\medskip

Finally, let us mention that
the numerical integration is performed with
dimensionless variables introduced 
by using natural units set by $\mu$, with
\begin{eqnarray} 
(r,z)\to (r,z)/\mu,~~{\rm and}~~
\omega \to \omega/\mu,
\end{eqnarray} 
(as well as $L \to L/\mu$).
As a result, the dependence on $\mu$ disappears from the
equations,
and the only input parameter is the (scaled)
frequency $\omega$ that enters the scalar field ansatz.
Also, the  quantities of interest are expressed in units set by $\mu$ and $G_5$. In order to simplify the output, however, we set
$G_5 = 1$ in all figures.
For the localized solitons with $\mu=0$ in Section (\ref{EKGlocalized}),
we shall use instead a scaling with respect to the 
length $L$ of the extra-dimension.
 
\subsection{The  uniform strings and the zero mode}
The uniform EKG strings were reported in a more general
context in the recent work \cite{Brihaye:2023vox}. 
They can be studied within the following ansatz 
\begin{eqnarray}
	\label{metricKK}
	ds^2= -b(r) dt^2+
	\frac{dr^2}{f(r)}+ r^2 d\Omega_2^2
	+ a(r)dz^2~,~~{\rm and}~~\Psi=\psi(r)e^{-i\omega t},
\end{eqnarray}
in terms of three metric functions and a scalar field (real) amplitude.
When performing a  KK  reduction
$w.r.t.$ the $z-$direction, these strings become 
$D= 4$ spherical BSs in the EKG-dilaton model
(\ref{action4}), with a vanishing $U(1)$ field. 

Unsurprisingly,
the properties of the uniform EKG strings are akin to
those of the well-known $D=4$ BSs.
For example,  
they also exist
 for $\omega<\mu$, which is a bound state
condition, with the global charges vanishing in that limit.
As we decrease the frequency, the mass increases until a maximum value. 
Continuing to decrease $\omega$, one finds a minimal frequency value
below which no solutions are found.
The $(w,M)$-curve then spirals towards a central region of the diagram\footnote{The same behaviour is found for the Noether charge.}, see
the corresponding curves in
Figure \ref{phases} (right panel) and  Figure \ref{wM1}.
Here we consider fundamental
solutions only, with a scalar profile 
interpolating
monotonically between a nonzero value at $r=0$
and $\psi=0$ at infinity.

In addressing the question of stability of the uniform solution, 
we have found convenient to  consider an ansatz with
\begin{equation}
\label{nuans}
ds^2 =-b(r) e^{2A(r,z)}dt^2 + e^{2B(r,z)}\left( \frac{dr^2}{f(r)} + a(r)dz^2 \right) + r^2e^{2C(r,z)} d\Omega_{2}^2,~{\rm and}~\Psi=\psi(r,z)e^{-i\omega t},~~{~~}
\end{equation}
where the non-uniformity is 
set in through the functions $A,B,C$. 
The uniform solutions discussed above are recovered for $A=B=C=0$.

Then,  following \cite{Gregory:1993vy} we perform an expansion of the functions 
$A,B,C,\psi$
 in terms 
of a small parameter $\epsilon$ 
and consider a Fourier series in the $z$-coordinate with
\begin{eqnarray}
\label{Xseries}
X(r,z) = \epsilon X_1(r) \cos(k z) + \Ord{\epsilon}{2},~
\psi(r,z) =\psi_0(r)+ \epsilon \psi_1(r) \cos(k z) + \Ord{\epsilon}{2},~
\end{eqnarray}
$X$ denoting generically $\{A,B,C \}$ and $k$ being the critical wavenumber corresponding to a static perturbation,
$k= 2 \pi/L$.
We then substitute the form 
\eqref{nuans} in the general EKG equations (\ref{EKG-eqs}) and  
expand the functions according to \eqref{Xseries}.
The equations for $\{A_1,B_1,C_1;\psi_1 \}$
are found by considering the linear order terms there in the infinitesimal parameter 
$\epsilon$. 
The system obtained in this way is  relevant for addressing the zero-mode problem. 
The $(r,z)$-component of the Einstein equations allows to eliminate the function
 $B_1$ in favor of the $A_1$ and $C_1$
 and to reduce the problem to a system of
 three ordinary differential equations for $A_1$, $C_1$ and $\psi_1$.
 These equations are 
rather complicated and not so enlightening,
being 
 solved numerically by employing  a multi-shooting
procedure.

\begin{figure}[h!]
\begin{center}
\includegraphics[width=0.7\textwidth]{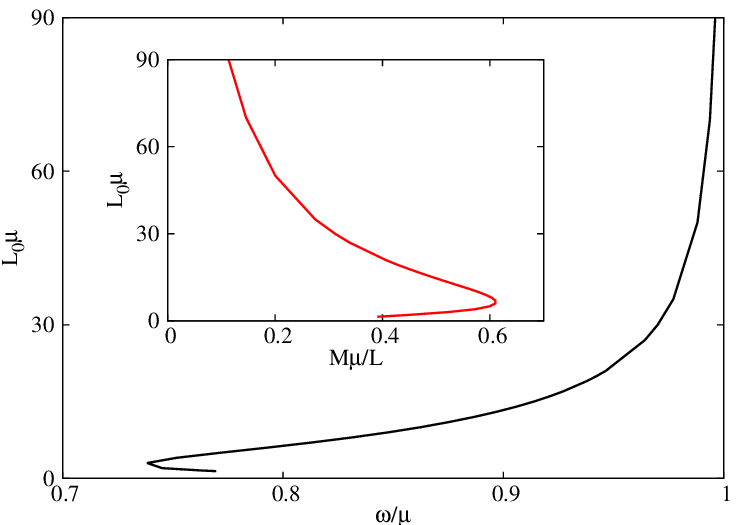} 
\caption{  
The critical length $L_0$
of the extra-dimension allowing for zero modes around the uniform EKG strings
is shown as a function of the scalar field frequency $\omega$,
and as a function of the solutions' mass (inset).
}
\label{GL1}
\end{center}
\end{figure}

For a given uniform string, 
as specified $e.g.$ by the input parameter $\omega$,
solutions with the right asymptotics
($e.g.$ $A_1$, $C_1$, $\psi_1$ vanish at infinity) 
can be accommodated only if the parameter $k^2$ assumes a particular value, which defines a critical length
$L_0$
of the extra-dimension which allows for a zero mode.
The numerical results  are shown in 
Figure \ref{GL1}, 
with some new features as compared to $Q$-strings.
As one can see, the critical length of the extra-dimensions
increases as $\omega\to \mu$
and  likely diverges again in that limit.
At the same time, $L_0$
decreases along the $M(\omega)$-curve, 
and
we conjecture that $L_0 \to 0$
for solutions
close to central inspiralling region of the BSs curve.

\subsection{Solutions with 
a dependence along the extra-dimension}

\subsubsection{The approach and quantities of interest}

Given the existence of zero modes together with the results for $Q$-strings,
it is clear that non-uniform configurations
should also exist.
In principle, these solutions can be
studied by employing the Ansatz 
(\ref{nuans}). 
However,  it turns out that
for this choice 
the numerical convergence was
typically very slow, while the 
constraint equations were not satisfied with enough 
accuracy\footnote{
A similar  situation has been found
when considering a different metric gauge choice,
with a conformally flat form of the $(r,z)$-part.}.
 
To overcome this problem,
we have chosen to
construct the $z$-dependent  solutions  
by employing the 
Einstein-De Turck  (EDT) approach, as proposed in~Refs. \cite{Headrick:2009pv,Adam:2011dn}, 
which
 has the advantage of not fixing \textit{a priori}  a metric gauge, 
yielding at the same time elliptic equations
(see~\cite{Wiseman:2011by,Dias:2015nua} for reviews).    
In this approach,  one solves a modified version
of the Einstein equations, with
\begin{eqnarray}
\label{EDT}
R_{ab}-\nabla_{(a}\xi_{b)}=
2 \left(T_{ab}-\frac{1}{2}T  g_{ab}\right) \ .
\end{eqnarray}
Here, $\xi^a$ is a vector defined as
$
\xi^a \equiv g^{bc}(\Gamma_{bc}^a
-\bar \Gamma_{bc}^a)\ ,
$
where 
$\Gamma_{bc}^a$ is the Levi-Civita connection associated to the
spacetime metric $g$ that one wants to determine, and a reference metric $\bar g$ is introduced, 
with $\bar \Gamma_{bc}^a$ being the corresponding Levi-Civita connection.
Solutions to (\ref{EDT}) solve the Einstein equations
iff $\xi^a \equiv 0$ everywhere.
To achieve this,
we impose boundary conditions  which are compatible with
$\xi^a = 0$
on the boundary of the domain of integration.
Then, this should imply $\xi^a \equiv 0$ everywhere,
a condition which is verified from  the numerical output. 

In this approach, we use  a metric Ansatz
with five functions, $\{A,B_1,B_2,C,S \}$
which depend on $(r,z)$: 
\begin{eqnarray}
ds^2=-e^{2A(r,z)}dt^2+e^{2B_1(r,z)}dr^2+e^{2B_2(r,z)}(dz+S(r,z) dr)^2
+e^{2C(r,z)}r^2  d\Omega_2^2 ,
{~~}
\end{eqnarray}
while the ansatz for the
scalar field is still given by  (\ref{psin}).
The problem reduces to
solve a set of six partial differential equations with suitable
boundary conditions (BCs).
The reference metric $\bar g$ 
corresponds to the ${\cal M}_4\times S^1$ background
($i.e.$  
$A=B_1=B_2=C=S=0$).
The BCs are found by 
 constructing an approximate form of the solutions on the
boundary of the domain of integration compatible with the requirement $\xi^a = 0$,
regularity  of the solutions plus the assumption of standard KK asymptoticss.
The imposed BCs read 
\begin{eqnarray}
&&
\partial_r A \big |_{r=0}=
\partial_r B_1 \big |_{r=0}=
\partial_r B_2 \big |_{r=0}=
\partial_r \psi \big |_{r=0}=0,~~
(C-B_1) \big |_{r=0}=0,~~
S\big |_{r=0}=0,
\\
\nonumber
&&
A \big |_{r=\infty}=
B_1 \big |_{r=\infty}=
B_2 \big |_{r=\infty}=
C \big |_{r=\infty}=
S \big |_{r=\infty}=
\psi \big |_{r=\infty}=0,~
\\
\nonumber
&&
\partial_z A\big |_{z=0,L}= 
\partial_z B_1\big |_{z=0,L}= 
\partial_z B_2\big |_{z=0,L}= 
\partial_z C\big |_{z=0,L}= 
S\big |_{z=0,L}= 
\partial_z \psi \big |_{z=0,L}=0  \ .
\end{eqnarray}
In addition, we
assume reflection symmetry $w.r.t.$ $z=L/2$,

The equation for $\{A,B_1,B_2,C,S; \psi \}$ are discretized in an non-equidistant 
$(r, z)$-grid, with
(usually) around 300×50 points. 
Using the Newton-Raphson approach, the resulting system is
solved iteratively until convergence is achieved, 
by employing a professional solver which uses a finite difference method
\cite{SCHONAUER1989279,SCHONAUER1990279,SCHONAUER2001473}. 
The typical numerical errors for the reported solutions
are estimated 
to be of the order of $10^{-4}$.


\medskip

Turning to quantities of interest,
a generic static spacetime with
${\cal M}_{4}\times S^1$ asymptotics
possesses two global charges,
the mass $M$ and the  tension ${\mathcal T}$,
which
are encoded in the asymptotics of the metric potentials
$g_{tt}$ and $g_{zz}$,
with
\begin{eqnarray}
\label{o1} 
g_{tt}\simeq -1+\frac{c_t}{r },~~~g_{zz}\simeq 1+\frac{c_z}{r},
\end{eqnarray}
$c_t$ and $c_z$ being two constants.
The results in 
\cite{Harmark:2003dg,Harmark:2004ch,Kol:2003if} 
implies that 
\begin{eqnarray}
\label{o2} 
M=\frac{\Omega_{2}L}{16 \pi G_5}(2c_t-c_z),
~~{\mathcal T}=\frac{\Omega_{2}}{16 \pi G_5}(c_t-2 c_z),
\end{eqnarray}
where $\Omega_2=4\pi$ is the area of the unit $S^2$-sphere.

The EKG string solutions posses also a conserved Noether charge
\begin{eqnarray}
\label{Q2} 
Q_N = 8\pi \omega \int_0^\infty
dr \int_0^Ldz ~e^{-A+B_1+B_2+2C}r^2 \psi^2.
\end{eqnarray}
Additionally, one can show that an horizonless configuration
satisfies the following Smarr relation:
\begin{eqnarray}
\label{Smarr} 
M =\frac{1}{2}{\mathcal T} L+{\cal U}_{(\Psi)},~~
{\rm with}~~{\cal U}_{(\Psi)}=\frac{1}{2}\int d^4x \sqrt{-g} (T-3T_t^t).
\end{eqnarray}
The following differential law also holds
\begin{eqnarray}
\label{1st} 
dM= \omega dQ+{\cal T}dL~.
\end{eqnarray}

\subsubsection{The non-uniform EKG strings}
\begin{figure}[h]
	\makebox[\linewidth][c]{%
		\begin{subfigure}[b]{10cm}
			\centering
\includegraphics[width=10cm]{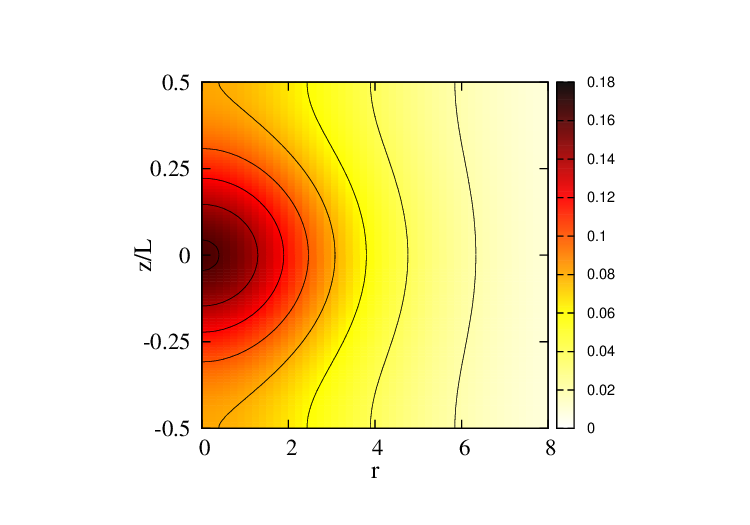}
		\end{subfigure}%
		\begin{subfigure}[b]{10cm}
			\centering
\includegraphics[width=10cm]{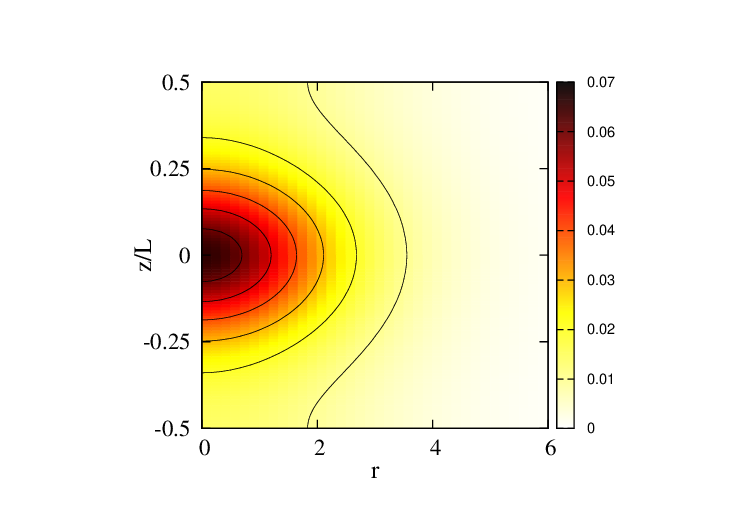}
		\end{subfigure}%
	}\\
	\caption{
	{\small
The amplitude of the scalar field (left) 
and the energy density (right) 
are shown as colour map for a typical
EKG 
non-uniform string.
The input parameters here are 
$L\mu=12$
and
$\omega/\mu=0.889$.
}
		\label{ZE}}
\end{figure}

%
As with the (non-gravitating) $Q$-strings,
given a value of $L$,
a set of $z-$dependent solutions emerges as non-linear continuation of the zero-mode 
of the corresponding uniform configuration.
The scalar amplitude and the energy density
of a typical solution
are shown as a colour map in Figure \ref{ZE}.
They resemble the case of non-uniform
$Q$-strings, the solution being again
localized on the $z-$circle.

%

In a $M(\omega)$-diagram, the non-uniform set possess an inspiraling 
behaviour towards a central configuration, with several different branches of
solutions, see the right panel in
Figure \ref{phases}. However, the numerics
become challenging along the secondary
branches, with a lower numerical accuracy.

\begin{figure}[h]
	\makebox[\linewidth][c]{%
		\begin{subfigure}[b]{8cm}
			\centering
\includegraphics[width=8cm]{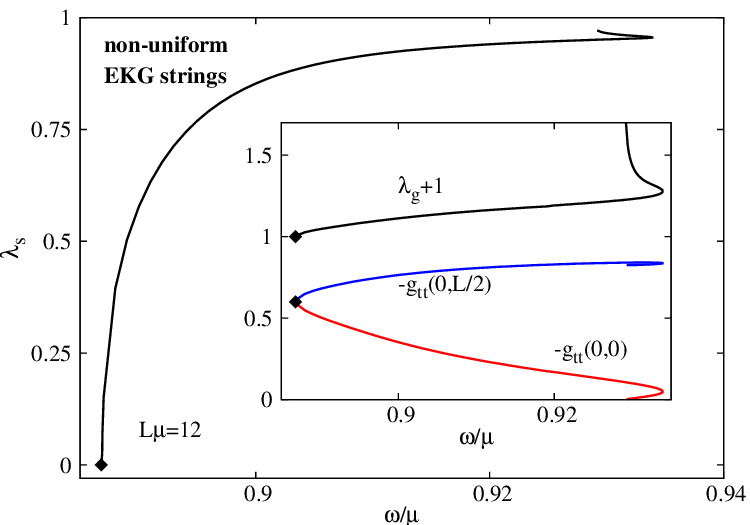}
		\end{subfigure}%
		\begin{subfigure}[b]{8cm}
			\centering
\includegraphics[width=8cm]{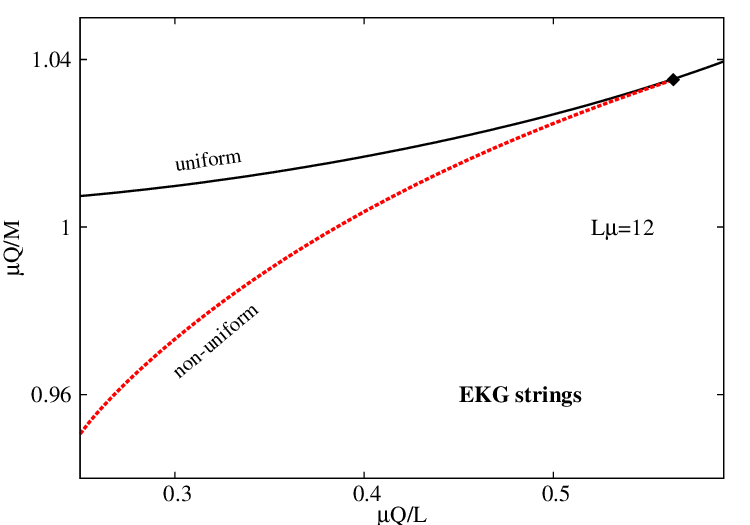}
		\end{subfigure}%
	}
 \\
	\caption{
{\small
Several features of a family of EKG strings with a fixed length of the extra-dimension.
The dots indicate a bifurcation with the non-uniform branch.  
}
		\label{features}}
\end{figure}

The deformation of the solutions increases along the spiral,
with a increasingly pronounced $z$-dependence of all functions.
 A simple geometrical measure of  the string deformation is provided by
\begin{eqnarray}
    \lambda_g=\frac{L_{max}^{(0)}}{L_{min}^{(0)}}-1,~~ 
\end{eqnarray}
where $L_{max}^{0}=e^{B_2(0,0)}$
($L_{min}^{0}=e^{B_2(0,L/2)}$)
is the maximal (minimal) proper size of the $z-$circle at $r=0$. 
This quantity monotonically increases from
 $0$ (the uniform limit),
and appears to increase without bounds along the spiral.
In other words,
the $z-$circle bulge out at some points and squeeze in at others.
Alternatively, one can take a deformation
measure in terms of the scalar field, 
\begin{eqnarray}
    \lambda_s=1-\frac{\psi(r=0,z=L/2)}{\psi(r=0,z=0)}~,
\end{eqnarray}
which again increases monotonically from  $0$ (the uniform limit),
likely reaching $1$ as the central configuration in the spiral is approached.
Also, let us mention that the 
metric function $g_{tt}(0,0)$
also increases monotonically along the branch of 
non-uniform solutions and likely vanishes at
the limiting central configuration.
These features are shown in 
Figure \ref{features}
(left panel).

\begin{figure}[h]
	\makebox[\linewidth][c]{%
		\begin{subfigure}[b]{8cm}
			\centering
\includegraphics[width=8cm]{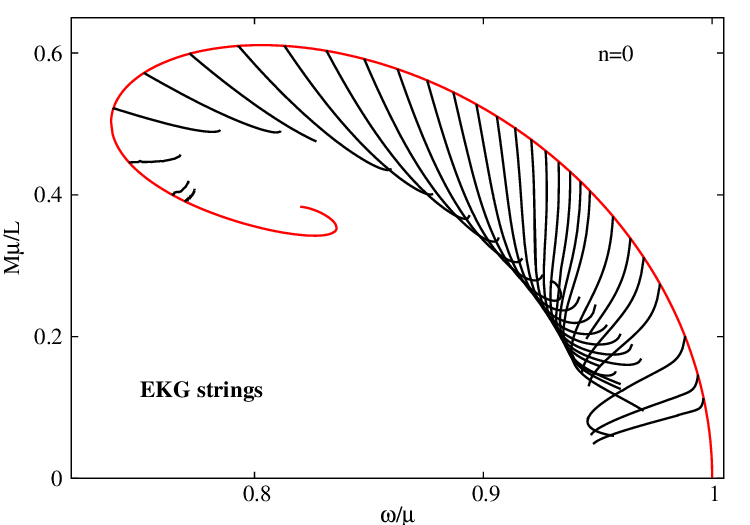}
		\end{subfigure}%
		\begin{subfigure}[b]{8cm}
			\centering
\includegraphics[width=8cm]{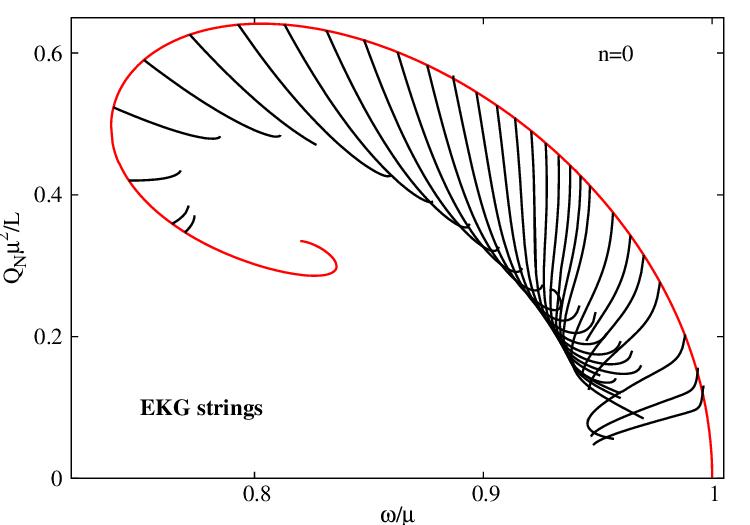}
		\end{subfigure}%
	}\\
	\caption{
	{\small
The mass $M$ and Noether charge $Q_N$ are shown for EKG 
uniform (red curve) and  non-uniform strings
with  different values of $L$ running from $L\mu=1.4$ 
(the last black curve on the lower branch)
to $L\mu=90$ (the black curve closest to the maximal frequency). 
}
		\label{wM1}}
\end{figure}

Figure  \ref{wM1}
provides an idea about the
domain of existence of solutions,
when varying the length $L$
of the extra-dimension. 
Various values of $L$ are considered there, running 
from $1.4$ to $90$. 
One can see that, for any branch,   the mass and the Noether charge of the non-uniform solutions  are  bounded
from above by the values of the corresponding uniform strings. 
Also, all non-uniform solutions have $\omega<\mu$,   although in principle
they could probe higher frequencies due to
the excited KK modes.

Another interesting feature
 is that, for a given Noether charge, 
 the mass is minimized by the uniform strings, 
 see the right panel in Figure \ref{features}.
 This suggests that the non-uniform 
 configurations cannot be the end point of the instability associated with the GL zero-mode\footnote{Note the (likely superficial)
 analogy with the situation found for 
 non-uniform black strings, which also are entropically disfavoured over the uniform ones.
 }.  

Finally, we mention that the 
index $n$ of the solutions,
as given by (\ref{n})
always vanishes; 
however,
one can conjecture that
the limiting configuration at the center of the spiral
would have
$\psi(r=0,z=L/2)=0$
and thus
$n=1$.
 
\subsubsection{The localized EKG solitons}
\label{EKGlocalized}
Analogous to the case of $Q$-solutions,
we have found numerical evidence 
for the existence of localized  solitons,
which again are supported
by the excited KK modes.
The scalar field solution (\ref{sup1})
is approached asymptotically,
the central singularity being regularized
 by the nonlinear gravity effects.
%
%
%
\begin{figure}[h]
	\makebox[\linewidth][c]{%
		\begin{subfigure}[b]{10cm}
			\centering
\includegraphics[width=10cm]{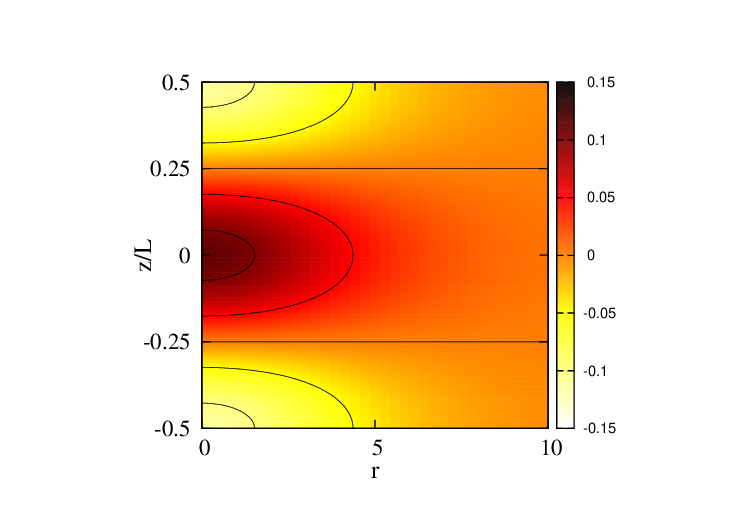}
		\end{subfigure}%
		\begin{subfigure}[b]{10cm}
			\centering
\includegraphics[width=10cm]{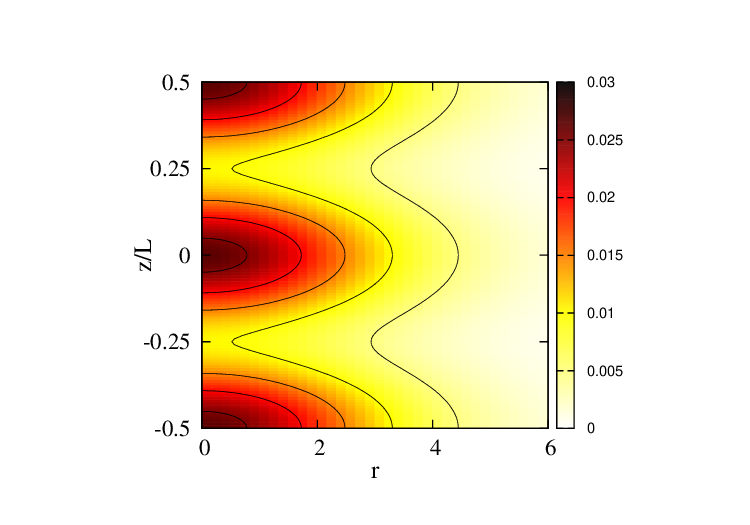}
		\end{subfigure}%
	}\\
	\caption{
	{\small
The amplitude of the scalar field (left) 
and the energy density (right) 
are shown as colour map for a typical
$n=2$
EKG  localized soliton.
The input parameters here are 
$L\mu=6$
and
$\omega.\mu=0.9$.
}
		\label{ZE-caged}}
\end{figure}
%
The configurations considered so far always possess 
two components localized on $z-$circle, 
 with two nodes of the scalar field amplitude,
see
 Figure \ref{ZE-caged}.

The mass-frequency diagram for
 localized EKG solitons
with $L\mu=12$  is shown in 
Figure \ref{phases} (right panel). 
The picture there is likely to be generic,
and we have constructed families of localized solitons
for few more values of the
(dimensionless) parameter $L\mu$.
Therefore such configurations are likely to exist for any value of 
$L$ 
and  we conjecture that the uniform set is approached as $L\to  \infty$.

 One can notice that localized EKG solitons form a  disconnected set, 
 with a  $M(\omega)$-diagram
which is rather similar with that found for 
the uniform strings.
 For example, the  solutions trivialize as 
$\omega \to \mu_{eff}= \sqrt{\mu^2+(\frac{2\pi}{L})^2}$.
Also, there are 
 more branches of solutions,
the $M(\omega)$ curve 
likely
describing 
a spiral twards a central configuration.

\begin{figure}[h!]
\begin{center}
\includegraphics[width=0.7\textwidth]{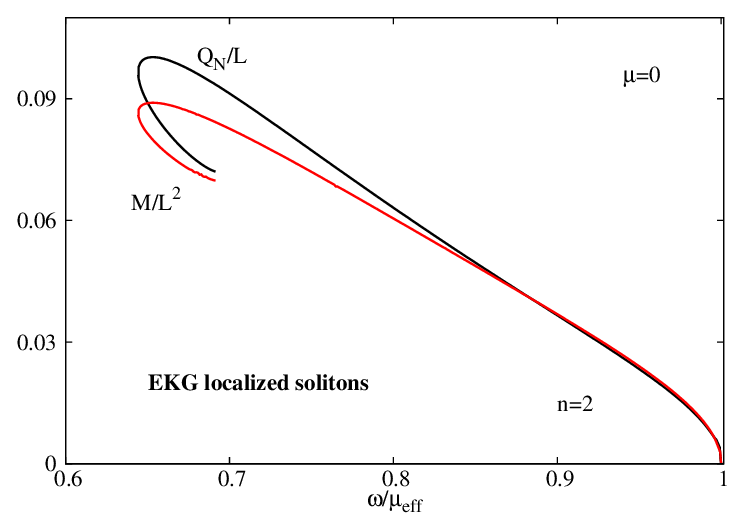} 
\caption{ 
The mass and the Noether charge are shown as a function of frequency for localized
EKG solitons with a vanishing field mass.
}
\label{wM-mu0}
\end{center}
\end{figure}

In addition, there are again solutions with
a massless scalar field, $\mu=0$,
the KK modes supplying an effective mass for the field and thus providing a confining mechanism ($i.e.$ no waves at infinity).
The  Smarr relation simplifies in this case, 
being a linear combination of solutions' mass, tension and Noether charge,
\begin{eqnarray}
\label{Smarr1} 
M =\frac{1}{2}({\mathcal T} L+3\omega Q_N) \ .
\end{eqnarray}
The  properties of these  solutions 
with $\mu=0$
are
qualitatively similar to
those found for a massive field.
For example,
the mass (and Noether charge) 
still exhibit an inspiralling behaviour in term of
the field frequency, see Figure \ref{wM-mu0}.
Also, the solutions with $\mu=0$ exist for any value of $L$, which in this case
provides the only intrinsic length  scale.

Finally, we mention that
for all localized solitons 
considered so far,
the number of nodes of the scalar field amplitude
in the $z$-direction is  $n=2$.

 \section{Conclusions and remarks}
 \label{final}

The study of configurations in more than 
$D = 4$ dimensions is a subject of long
standing interest in General Relativity, the case of a
KK theory with only one (compact)
extra-dimension providing the simplest model.
In a surprising development,
Gregory and Laflamme showed that the Schwarzschild black string solution is unstable
to linearized perturbations with a long wavelength along the circle
\cite{Gregory:1993vy}.
 This leads to the existence of a branch of
 (vacuum)
 solutions with a non-trivial dependence 
 of the extra-dimension, the {\it non-uniform} black strings.
 In addition, there are also {\it localized}
 black holes, which
 are Schwarzschild-Tangherlini--like near the horizon,
 despite approaching asymptotically the KK background.

 \medskip

Until recently, studies of the GL-type instability
(together with the emerging non-uniform configurations)
dealt with the case of black objects. Recently, however, a study of the GL instability for the case of $D=2+1+1$ flat spacetime $Q$-strings emerged~\cite{Chen:2024axd}, motivating the study of the consequent phases of equilibrium solutions. 

This work is a first attempt in this direction, restricting to the simplest case of a massive
complex scalar field.
The main reported results can be summarize as follows.
The uniform $Q$-strings are unstable to linearized perturbations, with a long wavelength along the circle. At the threshold of the instability, 
both the $Q$-strings and the EKG boson strings 
possess a GL-like zero-mode, with a non-trivial dependence of the extra-dimension. This supports the idea that the existence of
a GL-type instability
is a more general feature of configurations
extending into
an extra-dimension,  
not necessarily connected
with the presence of an horizon (with the associated thermodynamics).

The non-linear continuation of the zero-modes
result in a new phase of the $D=5$ solutions -- 
tbe {\it non-uniform} strings.  In addition, there are also intrinsic
 five-dimensional solutions -- 
 the {\it localized} solitons,
 which are supported by the KK (excited) modes of
 the scalar field. Note that such configurations exist as well for a
 massless scalar field.

 A list of possible directions for future work naturally emerges. One could find the unstable mode of the uniform EKG strings
and show that the solutions with $L>L_0$
are unstable, as expected. It would be interesting 
to clarify the connection with the Jeans
\cite{Harmark:2007md}
 and Rayleigh-Plateau
\cite{Cardoso:2006ks}  instabilities and to 
explain why the number of nodes $n$ in the $z$-direction
is conserved along sequences of solutions.
For example, why the non-uniform strings
always have $n=0$
and do not develop any nodes?

Apart from the fundamental solutions
in this work, one could also consider their excitations, for
which $\psi$ oscillates around zero
along the radial direction. 
These can be parameterized by the number 
of radial nodes of the scalar amplitude. 
Constructing the generalizations of the
localized solutions in a model with quartic self-interaction
($i.e.$ and extra $\lambda |\Psi|^4$ term in the action (\ref{action})) should also be possible.
Likely, as with the $D=4$ case \cite{Colpi:1986ye},
the  resulting configuration may differ markedly from
the non-interacting case, with a much larger mass,
$M_{max}\sim \sqrt{\lambda}$.
Technically more difficult, one could try to find the  spinning generalizations of the solutions 
in this work,
together with their black holes counterparts.

Let us conclude with two more general questions.
Firstly, for all unstable configurations what
the end-state of the perturbed system?
Do the non-uniform configurations occur dynamically 
(when perturbing a uniform solution), being the end point of evolution?
And what is the status of the localized solitons?
We recall that, for the same Noether charge, the 
non-uniform EKG strings are more massive than
the uniform strings, and so could not be the putative new
end-state.
And what is the situation with $Q$-solutions?
Secondly,
similar features should appear in models with a more realistic matter content.
For example, we predict the existence of non-uniform 
non-Abelian magnetic vortices 
for the Einstein--Yang-Mills-Higgs
model in
Ref. \cite{Volkov:2001tb}.

\section*{Acknowledgements}
The authors thank Pedro Cunha for useful discussions. 
This work is supported by the Center for Research and Development in Mathematics and Applications (CIDMA) through the Portuguese Foundation for Science and Technology (FCT -- Fundaç\~ao para a Ci\^encia e a Tecnologia) under the Multi-Annual Financing Program for R\&D Units, PTDC/FIS-AST/3041/2020 (\url{http://doi.org/10.54499/PTDC/FIS-AST/3041/2020}),  2022.04560.PTDC (\url{https://doi.org/10.54499/2022.04560.PTDC}) and 2024.05617.CERN (\url{https://doi.org/10.54499/2024.05617.CERN}). This work has further been supported by the European Horizon Europe staff exchange (SE) programme HORIZON-MSCA-2021-SE-01 Grant No.\ NewFunFiCO-101086251.

\bibliographystyle{JHEP}
\bibliography{main}

\end{document}